\documentclass[letterpaper,12pt]{article}
\usepackage{osajnl}
\begin{document}
\title{Detection of phase singularities with a Shack-Hartmann wavefront sensor}

\author{Mingzhou Chen, Filippus S. Roux, and Jan C. Olivier}

\address{Department of Electrical, Electronic and Computer Engineering, University of Pretoria,\\Lynnwood road, Pretoria, 0002, South Africa}

\maketitle

\begin{abstract}
While adaptive optical systems are able to remove moderate wavefront distortions in scintillated optical beams, phase singularities that appear in strongly scintillated beams can severely degrade the performance of such an adaptive optical system. Therefore, the detection of these phase singularities is an important aspect of strong scintillation adaptive optics. We investigate the detection of phase singularities with the aid of a Shack-Hartmann wavefront sensor and show that, in spite of some systematical deficiencies inherent to the Shack-Hartmann wavefront sensor, it can be used for the reliable detection of phase singularities, irrespective of their morphologies. We provide full analytical results, together with numerical simulations of the detection process.
\end{abstract}


\section{Introduction}

The Shack-Hartmann wavefront sensor (SHWS) is a widely used system to measure the shape of the wavefront of an optical beam that was scintillated after propagating through a turbulent atmosphere. The reasonable simplicity of the SHWS makes it ideal for use in an adaptive optical system, which is to correct the wavefront distortions in such a scintillated beam.

Various techniques exist\cite{Hudgin77, Fried77, Southwell80, Herrmann80, Dam02, Talmi06} to compute the shape of the wavefront from the raw data received as output from a SHWS. However, in strongly scintillated beams these techniques tend to fail\cite{Barchers02} due to the presence of phase singularities, which represent optical vortices \cite{Nye74, Coullet89} in the propagating scintillated beam. The failure of these techniques is to some extent a result of the apparent insensitivity of the SHWS to the phase function of a phase singularity,\cite{Fried98} which led to it being referred to as hidden phase. Phase singularities are points where the phase is undefined and where the phase around the singularity goes continuously through all phase values from 0 to $2\pi$. The direction (right-handed or left-handed) of increasing value indicates the topological charge ($\pm 1$) of the phase singularity. The phase function around a singularity can in general be anisotropically scaled along a particular direction, which gives rise to its morphology. An isotropic singularity is referred to as being canonical and anisotropic singularities are called noncanonical.

Initial least-squares phase reconstruction methods,\cite{Hudgin77, Fried77, Herrmann80, Southwell80} that were used in adaptive optical sytems, were based on a model for the phase slopes that makes the assumption that the phase function is continuous. The existence of phase singularities violates this assumption. Therefore, these methods can only be applied in application where the phase distortions are not large enough to generate phase singularities.

More recently, various authors\cite{Arrasmith99, Tyler00, Aksenov02,Hattori03} considered the separation of the phase gradient function into a continuous (curl-free) part and a solenoidal or rotational part. The former can be used in the adaptive optical system to correct the continuous part of the phase distortions. The latter is generally treated as a noise term, which is discarded. Since it represents the contribution of the phase singularities, the solenoidal part of the phase distortions cannot be removed in the same way that the continuous part of the phase distortions is removed. Neither does it simply go away after the continuous part has been corrected. It is therefore important to consider this solenoidal part of the phase distortions more carefully in order to find a way to get rid of the phase singularities. In this paper we consider the detection of the locations and topological charges of the phase singularities with the aid of the solenoidal part of the phase gradient function. We show that the SHWS is not completely insensitive to the phase of a singularity. In other words, this phase is not completely hidden, but is contained in the solenoidal part of the output of the SHWS. One can therefore use it to locate the phase singularities that are present in the wavefront.

The detection of the phase singularities is significantly influenced by the averaging effects of the SHWS subapertures. This averaging effect has not to date received much attention. Herrmann\cite{Herrmann81} mentioned the model error due to the averaging gradients, but did not give a detailed analysis. Aksenov\cite{Aksenov02} calculated the average wavefront slopes with a regularization procedure that produces continuous infinitely differentiable functions, which is not true when there are singularities in the phase function. In this paper, we analyze the effect of the averaging process on a phase singularity, by performing the actual integration of the phase gradient function of such a singularity over the area of a subaperture. From this result one can then show that it is possible to extract the required information of the phase singularities regardless of the effects of the averaging process. For a continuous vector field the singularities can be identified by computing the curl of this vector field. The output from a SHWS is in the form of a sampled vector field that approximately represents the gradient of the phase function of the scintillated beam. We implement the curl operation that operates on this sampled vector field in terms of finite differences. This finite difference curl operation is referred to as the circulation of the sampled vector field. The complete detection process thus consists of the averaging process performed by the SHWS and the circulation process performed on its output. The effect of the morphology and relative position of the singularity on this detection process is also investigated.

The remainder of the paper is organized as follows. In Section \ref{basics} we'll discuss the basic principle upon which the detection of phase singularities is based and provide some notation. The operation of the SHWS is discussed in Section \ref{shack}. In Section \ref{detect} we discuss the detection of phase singularities with the output of the SHWS. We analyse this for a canonical phase singularity and compute the peak value in Subsection \ref{canon} and then repeat this analysis for the more general noncanonical phase singularity in Subsection \ref{noncan}. The complete expression of the circulation for a noncanonical phase singularity is given in Appendix \ref{alg}. A numerical simulation is provided in Section \ref{num} to test the procedure on more realistic data. A summary and conclusions are provided in Section \ref{concl}. 

\section{Basic principle of operation}
\label{basics}

The phase function of a scintillated optical beam can in general be represented by the sum of a continuous phase function and an arbitrary number of phase singularities,
\begin{equation}
\theta(x,y) = \theta_{\rm C}(x,y) + \sum_n \phi(x-x_n,y-y_n;\alpha_n,\beta_n) ,
\label{fase}
\end{equation}
where $\theta_C(x,y)$ is the continuous phase function and $\phi(x-x_n,y-y_n;\alpha_n,\beta_n)$ represents the phase function of a noncanonical singularity, located at $(x_n,y_n)$. The latter phase function is given by,
\begin{equation}
\phi(x,y;\alpha,\beta) = -{i\over 2} \ln \left[ {\xi (x+iy)+\zeta (x-iy) \over \xi^* (x-iy)+\zeta^* (x+iy)} \right] ,
\label{vortnph}
\end{equation}
where $^*$ represents the complex conjugate and the morphology of the singularity is parameterized by,
\begin{eqnarray}
\xi & = & \cos(\alpha/2) \exp(i\beta/2) , \nonumber \\
\zeta & = & \sin(\alpha/2) \exp(-i\beta/2) , 
\label{morfz}
\end{eqnarray}
in terms of the morphology angles $0\leq\alpha\leq\pi$ and $0\leq\beta<2\pi$.

The presence of a singularities in a phase function can be determined with the aid of a closed line integral over the gradient of the phase function,
\begin{equation}
\oint_C \nabla\theta\cdot{\rm d}l = \tau 2\pi , 
\label{lineint}
\end{equation}
where $C$ denotes the closed integration contour; $\nabla\theta$ is the gradient of the phase function and $\tau$ is an integer that presents the net topological charge of all the singularities enclosed by the contour. Unfortunately the closed line integral in Eq.~(\ref{lineint}) is not convenient for the practical implementation of singularity detection.

To implement a singularity detection procedure with a physical system one can use the output of a SHWS, which measures the gradient of the phase function. For the moment we'll assume that this measurement is ideal and we'll return to the effect of the finite subapertures of the SHWS in Section \ref{shack}. One can view the output of a SHWS as a vector field ${\bf G} = G_x(x,y) \hat{x} + G_y(x,y) \hat{y}$. If this vector field represents the gradient of a continuous phase function, we must have,
\begin{equation}
{\partial G_x(x,y) \over \partial y} = {\partial G_y(x,y) \over \partial x} .
\label{restrict}
\end{equation}
If one finds that this is not the case then it indicates that ${\bf G}$ cannot be the gradient of a continuous phase function. This points to the presence of phase singularities. The extent to which the equality in Eq.~(\ref{restrict}) fails is given by the difference between its left-hand and right-hand sides, which becomes the curl of the vector field. Since the vector field only has $x$- and $y$-components that only depend on $x$ and $y$, its curl is a scalar which represents the $z$-component given by
\begin{equation}
D(x,y) = \nabla_T \times {\bf G}(x,y) .
\label{distrib}
\end{equation}

If the phase function of a scintillated optical beam is continuous (without phase singularities), the curl of the gradient of the phase function would give zero. On the other hand, if there are phase singularities in the phase function of the beam then the curl of this gradient is not zero.\cite{Roux06} Each phase singularity gives a Dirac delta function at the location of the singularity,
\begin{equation}
\nabla\!\!\times\!\!\nabla\phi(x,y,\alpha,\beta) = \tau
2\pi\delta(x)\delta(y) , 
\label{dirac}
\end{equation}
where $\tau(=\pm 1)$ is the topological charge of the phase singularity. So the curl of the gradient of a phase function $\theta(x,y)$ that contains phase singularities gives a sum of Dirac delta functions, each multiplied with the topological charge of the singularities,
\begin{equation}
\nabla\!\!\times\!\!\nabla\theta(x,y) = 2\pi\sum_n \tau_n
\delta(x-x_n) \delta(y-y_n) . 
\label{diracsom}
\end{equation}
By the same token the curl of the output vector field of a SHWS, shown in Eq.~(\ref{distrib}), gives us, according to Eq.~(\ref{diracsom}), a distribution of singularities, showing us where they are located and what their topological charges are. One can easily compute $D(x,y)$, as shown in Eq.~(\ref{distrib}), from ${\bf G}$ and then use the result to identify the phase singularities in the wavefront.

In the above discussion we assumed the ideal case, which ignores the effects of the discreteness and finite size of the subapertures of the SHWS. In the following section we consider the SHWS more carefully. We'll see that the subaperture size sets a scale beneath which singularities become unobservable and it also modifies the weight of $2\pi$ with which each singularities is multiplied.

\section{Shack-Hartmann wavefront sensor}
\label{shack}

The SHWS uses a lenslet array to sample the wavefront, as shown in Fig.~\ref{dia}. The slope of the wavefront for each of these samples is given by the location of the focal point formed by each lenslet in its back focal plane. To simplify our analysis we'll assume that the lenslet array is a square array of lenslets, each with a square shaped subaperture, as shown in Fig.~\ref{array}. A detector array is placed in the back focal plane of the lenslet array. A small subarray of this detector array is dedicated to each lenslet of the lenslet array. This subarray is used to determine the location of the focal point produced by each lenslet. The location of the focal point for the $(m,n)$-th subaperture (lenslet) is given by a position vector ${\bf u}^{m,n}$ that is given by the centroid of the measured intensity distribution in the back focal plane,
\begin{eqnarray}
{\bf u}^{m,n} = \frac{\int_H I({\bf u}){\bf u}\ {\rm d}^2u}{\int_H I({\bf u})\ {\rm d}^2u}-{\bf u_0}^{m,n},
\label{centroid}
\end{eqnarray}
where ${\bf u_0}^{m,n}$ would be the location of the focal point for a normally incident plane wave; $I(\bf u)$ is the intensity distribution over the detector plane; $H$ is the window of the subarray of detectors on the detector plane; and ${\rm d}^2u$ is the two-dimensional integration measure on the detector plane.

The average phase slope for each subaperture is given by the location of the focal point, as determined by the intensity centroid in Eq.~(\ref{centroid}). This relationship follows from the Fourier relationship, as portrayed in Fig.~\ref{ft}, between the phase tilt of the incident wave in front of a lens and the location of the resulting focal point behind it. So the average phase slope over the $(m,n)$-th subaperture can be expressed as
\begin{equation}
{\bf G}^{m,n} = {\int_\Omega \nabla \theta({\bf x})\ {\rm d}^2x \over \int_\Omega\ {\rm d}^2x} = \frac{k}{f} {\bf u}^{m,n},
\label{meanslope}
\end{equation}
where $\nabla \theta({\bf x})$ is the gradient of the phase function of the incident wave; ${\bf x}$ denotes the two-dimensional position vector on the lenslet plane; $k(=2\pi/\lambda)$ is the wave number; $f$ is the focal length of the lenslets; $\Omega$ is the area of the lenslet subaperture; and ${\bf u}^{m,n}$ is the location of the $(m,n)$-th focal point on the detector plane, as defined in Eq.~(\ref{centroid}). The ${\bf G}^{m,n}$-values given by Eq.~(\ref{meanslope}) represent the sampled output vector field of the SHWS. Each value is associated with a point in the center of the particular subaperture, as denoted by the dots in Fig.~\ref{array}.

\section{Detection of phase singularities}
\label{detect}

In the previous section we saw that the output of the SHWS in a practical adaptive optical system is a sampled vector field ${\bf G}$ with each sample representing the averaged phase slopes over one subaperture. The solenoidal part of this vector field contains the information about the locations of the singularities, as pointed out in Section \ref{basics}. To extract this information from the sampled vector field, one needs to implement the curl operation of Eq.~(\ref{distrib}) numerically. Such a numerical implementation is equivalent to the line integral of Eq.~(\ref{lineint}). The numerical computation is done by computing what we refer to as the circulation, given by,
\begin{eqnarray}
D^{m,n} & = & \frac{w}{2} \left( G_x^{m,n}+G_x^{m,n}+G_y^{m,n+1}+G_y^{m+1,n+1} \right. \nonumber\\
& & \left. -G_x^{m+1,n+1}-G_x^{m+1,n}-G_y^{m+1,n}-G_y^{m,n} \right),
\label{circ}
\end{eqnarray}
where $w$ is the subaperture window size and the superscript $m,n$ denotes the subaperture index. The physical implementation of the circulation process is presented in Fig.~\ref{foursub}. The samples used in the calculation are denoted by the points in the centers of the subapertures shown in Fig.~\ref{foursub}. The circulation represents a line integral performed over the four subapertures along a contour denoted by the dashed-line. Note that the result of this circulation operation represents a value $D$ that should be associated with a point in the center of the four subapertures. However, we are assigning it the same index $m,n$ that is associated with the upper left subaperture.

The result of this computation $D^{m,n}$ is a distribution of the topological charges of the phase singularities. It is positive (negative) at the locations of phase singularities with positive (negative) topological charges and it should be zero where there are no phase singularities. However, this distribution is affected by the averaging process, the sampled nature of the data and, of course, noise. As a result the values are not exactly zero when there are no singularities. Moreover, the value of $D^{m,n}$ at the location of a single positive singularity is not $2\pi$ as one would expect if the circulation is an exact implementation of the line integral Eq.~(\ref{lineint}). In the next two subsections, we analyze the effects of the practical implementation of the singularity detection procedure.

\subsection{Canonical phase singularities}
\label{canon}

Here we consider the situation where a canonical singularity is located in the center of the four subapertures, as shown by point {\bf A} in Fig.~\ref{foursub}. We define the origin $(0,0)$ of our coordinate system at this point. The complex amplitude function of a canonical phase singularity at the origin can be expressed as $(x\pm iy)/r = \exp(\pm i\phi)$, where the sign indicates the topological charge of the singularity and $r$ and $\phi$ are respectively the radial coordinate and the azimuthal coordinate. Here we consider a positively charged singularity. The phase function of the singularity is simply the azimuthal coordinate $\phi$, provided that the singularity is located at the origin. The gradient of the phase function of a canonical phase singularity can then be expressed in Cartesian coordinates, as,
\begin{equation}
\nabla\phi(x,y) = {x\hat{y}-y\hat{x}\over x^2+y^2} .
\label{cangrad}
\end{equation}
Substituting Eq.~(\ref{cangrad}) into Eq.~(\ref{meanslope}), we obtain an analytical expression for the average phase slope in the $(m,n)$-th subaperture, given by,
\begin{eqnarray}
{\bf G}^{m,n} & = & \frac{1}{w^2}\int_{-w}^{0}\int_{-w}^{0}\frac{x\hat{y}-y\hat{x}}{x^2+y^2}\ {\rm
d}x{\rm d}y \nonumber \\
& = & \left( \frac{\pi}{4w}+\frac{\ln2}{2w} \right)(\hat{x}-\hat{y}) .
\label{integ}
\end{eqnarray}
The integration boundaries are determined by the location of the $(m,n)$-th subaperture within the four subaperture area shown in Fig.~\ref{foursub}. Due to the symmetry of the phase function of a canonical singularity at the origin, the average phase slope values for the other subapertures will be the same apart from a possible change in sign. Then, according to Eqs.~(\ref{circ}) and (\ref{integ}), the value for $D^{m,n}$ will be,
\begin{equation}
D^{m,n} = 4w G_x^{m,n} = \pi+2\ln(2) = 4.527887 .
\label{cancirc}
\end{equation}
We note that the value of the circulation is $\pi+2\ln2$ and not the $2\pi$ that one finds for the analytical case given by Eq.~(\ref{lineint}). It is the averaging process that is responsible for this difference and not the finite differences of the calculation process. One finds that if the same finite difference circulation calculation, Eq.~(\ref{circ}), is performed on a sampled gradient function of a canonical phase singularity without the averaging process the result is indeed $2\pi$.

Although it is the averaging process that causes the difference, we note that this circulation value is independent of the subaperture window size $w$. This is because the phase function of a singularity is scale invariant. So the difference is produced simply because some averaging takes place, but it does not matter how big the subapertures are over which the slopes are averaged.

When the singularity is not located at the origin, in the center of the four apertures, but at some other location such as the position ${\bf B}$ at $(x_0,y_0)$ in Fig.~\ref{foursub}, we can expect the value of the circulation to change. The expression for the gradient of the phase function now becomes
\begin{equation}
\nabla\phi(x-x_0,y-y_0) = {(x-x_0)\hat{y}-(y-y_0)\hat{x}\over(x-x_0)^2+(y-y_0)^2} .
\label{skuifgrad}
\end{equation}
Using Eq.~(\ref{skuifgrad}) and Eq.~(\ref{meanslope}), one can compute the average phase slopes ${\bf G}$ for the four subapertures. Then, with the aid of Eq.~(\ref{circ}), one can compute $D^{m,n}$. The final expression is rather complicated. It can be obtained from the more general expression for arbitrary morphology provided in Appendix \ref{alg}, as explained there.

The circulation $D$ is shown in Fig.~\ref{canshape} as a function of the location of the singularity in terms of normalized coordinates $\mu=x_0/w$ and $\nu=y_0/w$. A topview of this function over the region $-2<\mu<2$ and $-2<\nu<2$ is shown in Fig.~\ref{canshape}(a). The precise shape of the circulation function as one-dimensional functions of $\mu$ are shown in Fig.~\ref{canshape}(b) for $\nu=0$ (middle line) and for $\nu=\mu$ (diagonal). At the origin $(\mu,\nu)=(0,0)$ the circulation function has a peak with the value $D=4.53$, consistent with Eq.~(\ref{cancirc}). Away from the origin the value of $D$ decreases rapidly and approaches zero for $|\mu|,|\nu|>1$. In some regions the value of $D$ drops below zero. At the corners of the four subaperture area, where $|\mu|,|\nu|=1$, the circulation function forms negatively valued peaks with the value,
\begin{equation}
D = {\pi\over 2}-2\ln(5)-2\arctan(2)+5\ln(2) = -0.396641 .
\label{corner}
\end{equation}
The fact that the circulation function never becomes exactly zero implies that the existence of a singularity at some point in the output plane gives nonzero circulation values in all other parts of the output plane.

In view of the fact that the peak value of the circulation function is $\pi+2\ln(2)$ instead of $2\pi$, it is interesting to note that, if one would add the four sample values of $D^{m,n}$ closest to the location of a singularity -- i.e.\ the four values that surround the singularity -- then the result is closer to $2\pi$.

\subsection{Noncanonical phase singularities}
\label{noncan}

Singularities in strongly scintillated beams are in general noncanonical. It is therefore necessary to know how the computation of the circulation is affected by the morphology of the singularity. The phase function of a singularity with an arbitrary morphology is given by Eqs.~(\ref{vortnph}) and (\ref{morfz}). The gradient of this phase function can be expressed by,
\begin{equation}
\nabla\theta(x,y) = {(x\hat{y}-y\hat{x})C\over x^2(1+A)-2yxB+y^2(1-A)} .
\label{noncangrad}
\end{equation}
where,
\begin{eqnarray}
A & = & \sin(\alpha) \cos(\beta) , \label{parma} \\
B & = & \sin(\alpha) \sin(\beta) , \label{parmb} \\
C & = & \cos(\alpha) . \label{parmc}
\end{eqnarray}
with the morphology angles $\alpha$ and $\beta$, as defined in Eq.~(\ref{morfz}). The singularity can be translated to any location $(x_0,y_0)$ by replacing $x \rightarrow x-x_0, y \rightarrow y-y_0$ in Eq.~(\ref{noncangrad}).

We use the same procedure that was use in Subsection \ref{canon} to analyze the noncanonical case. We use Eq.~(\ref{meanslope}) to compute the output of the SHWS for the phase gradient given in Eq.~(\ref{noncangrad}), shifted to $(x_0,y_0)$. Then we substitute the result into Eq.~(\ref{circ}) to compute the circulation $D$. In Appendix \ref{alg} we provide the complete expression of this circulation function for a singularity with an arbitrary morphology, located at an arbitrary point (in normalized coordinate $\mu=x_0/w$ and $\nu= y_0/w$) within the four subapertures, shown in Fig.~\ref{foursub}.

In Fig.~\ref{peak} we show the peak value at $(\mu,\nu)=(0,0)$ for $D$ as a function of the morphological angles $\alpha$ and $\beta$. Note that the peak is positive (negative) for $0\leq\alpha<\pi/2$ ($\pi/2<\alpha\leq\pi$). At $\alpha=\pi/2$ there is a discrete jump, which represents the change in the topological charge of the singularity. The fact that there is such a large difference between the values on either side of the jump, indicates that the circulation $D$, calculated from the output of SHWS, can in principle determine the topological charge of a singularity even if it is severely anisotropic. However, in a practical situation, singularities tend to have such severe anisotropic morphologies only when they appear in oppositely charged pairs close to each other. In such situations their circulation values will in general partially cancel each other, making them difficult to identify. Near the jump at $\alpha=\pi/2$ the value of $D$ has its greatest fluctuation as a function of $\beta$. Next to the jump $D$ fluctuates between $\pi$ and $2\pi$. When $\alpha$ approaches the canonical values of $0$ or $\pi$ the value of $D$ tend towards its canonical peak value of $4.53$ and the fluctuations as a function of $\beta$ diminish.

In Figs.~\ref{ex1} and \ref{ex2} we provide plots of $D$ over the region $-2<\mu<2$ and $-2<\nu<2$, for two different morphologies of the singularity. First we consider the case when $\alpha=\pi/4$ and $\beta=\pi$. This represents a singularity with a moderate anisotropy oriented along the $y$-axis. The topview of the circulation function for this case is shown in Fig.~\ref{ex1}(a) and the shape of the circulation function is shown in Fig.~\ref{ex1}(b) in terms of three one-dimensional functions:\ along $\nu=0$ ($\mu$-line), along $\mu=0$ ($\nu$-line) and along the line where $\nu=\mu$ (diagonal line). The peak value at the origin is about $4.5$. The function then decreases toward zero away from the origin. We note that the shape of the circulation function is more anisotropic than the shape in Fig.~\ref{canshape} in that the respective widths of the peak along the $\mu$- and $\nu$-directions are not equal. There are still regions where the function becomes negative.

Next we consider the case where $\alpha=4\pi/9$ and $\beta=\pi/2$. This represents a highly anisotropic singularity oriented diagonally along the line where $\nu=-\mu$. We show the topview of the circulation function for this case in Fig.~\ref{ex1}(a) and the shape of the circulation function in Fig.~\ref{ex1}(b) in terms of three one-dimensional functions:\ along $\nu=0$ (middle line), along the line where $\nu=\mu$ (diagonal-I line) and along the line where $\nu=-\mu$ (diagonal-II line). The peak value at the origin is now about $5$. The function decreases away from the origin, but the decrease is much slower along the orientation of the singularity. The shape of the circulation function is therefore much more anisotropic.

\section{Numerical simulation}
\label{num}

Here we present a numerical simulation to test the singularity detection procedure that was analytically investigated in the previous sections. We simulate the propagation of a Gaussian beam over a distance of 100~km through a turbulent atmosphere. We use the well known numerical method of Ref.~\onlinecite{Roggemann00a} to perform the simulation. The strength of the turbulence is parameterized with $C_n^2=4\times 10^{-18}$~m$^{2/3}$ and we use 10 equally spaced phase screens to simulate the 10~km thick turbulent layers. This method provides a reasonable agreement between real world data and the simulation data. The phase of the beam is distorted when it reaches the system aperture and in our example we find four phase singularities in the wavefront, as shown in Fig.~\ref{numfig}(a). This beam then passes through our simulated SHWS, from which we obtain the sampled vector field ${\bf G}$, computed with Eq.~(\ref{meanslope}). The circulation function $D$ is then computed with Eq.~(\ref{circ}). The resulting circulation function is shown in Fig.~\ref{numfig}(b). Two of the singularities, one positive and one negative, are easily identified from their respective circulation values of $2.95$ and $-3.54$, in the lower-left corner of Fig.~\ref{numfig}(a). Note that the magnitudes of both these values are smaller than $2\pi$. This would be due to a combination of the fact that the singularities have noncanonical morphologies, the fact they are not located at the ideal location in the center of four subapertures and noise that is present in the phase function. Integrating over a $3\times 3$ neighborhood around these singularities we obtain values of $6.20$ and $-5.86$, respectively, which are closer to $\pm 2\pi$.

The other two singularities in the upper-right corner of Fig.~\ref{numfig}(a) are much closer to each other. Therefore, their individual circulation functions overlap and, having opposite topological charges, they partially cancel each other. As a result the circulation peaks that represent these singularities are severely diminished. For example, the positive peak for this pair of singularities in Fig.~\ref{numfig}(b) has a value of only $1.41$. Oppositely charged singularities that are located closer to each other are therefore more difficult to detect.

\section{Conclusion}
\label{concl}

The phase gradient that is produced as output of a SHWS contains information about the continuous phase function of the incident wave, but also of the phase singularities in the wavefront. A least-squares projector\cite{Arrasmith99, Tyler00, Aksenov02, Hattori03} can be used to extract the information about the continuous phase and can be used to correct contious phase distortions. The information about the singularities is contained in the solenoidal part of the phase gradient. The curl of this part gives a topological charges distribution, which represents the locations and topological charges of the singularities. Theoretically each positive (negative) singularity should be indicated by a value of $2\pi$ ($-2\pi$) in the topological charge distribution.

The averaging process inherent to the SHWS has a significant effect on the computed topological charge distribution. Instead of the theoretical value of $2\pi$, the actual value that is produced is at most about $4.53$. The precise location of the singularity relative to the subapertures in the SHWS, as well as the morphology of the singularity produce further variations in the value of the topological charge distribution at the location of a singularity. Nevertheless, these values are generally large enough to identify an isolated singularity. It is therefore possible to extract the information of the location and topological charge of the singularities from the output obtained from a Shack-Hartmann wavefront sensor.

In the analytical investigation presented here we only considered one singularity. In the numerical simulation we found that when different oppositely charged singularities are in close proximity to each other, their respective topological charge distributions, as produced by the circulation computations, would overlap, causing partial cancelation and a subsequent reduction in their peaks. This makes detection of these singularities more difficult. Further investigation is needed to understand the effect of multiple singularities located near each other on the detection process. 

The phase functions of scintillated optical beams are in general noisy. The noise becomes larger as the scintillation increases. The analysis that is provided here does not specifically include such noise. It is expected that the ability to detect singularities would deteriorate as the noise is increased. This is an important aspect that still needs to be investigated.

\appendix

\section{Circulation function for a general noncanonical singularity at an arbitrary location}
\label{alg}

Here we provide the general expression obtained when we compute the circulation, Eq.~(\ref{circ}), of the sampled average slope ${\bf G}$, Eq.~(\ref{meanslope}), obtained as output from the SHWS when the input is the phase function of a noncanonical singularity, located at $\mu=x_0/w$ and $\nu=y_0/w$ with morphology angles $\alpha$ and $\beta$,
\begin{eqnarray}
D^{m,n} & = &
    {(\mu+1)(A_m-B) \over 2 A_m} \arctan \left[ {(\mu+1) B-(\nu-1) A_m\over (\mu+1) C} \right] \nonumber \\
&& -{(\mu+1)(A_m+B) \over 2 A_m} \arctan \left[ {(\mu+1) B-(\nu+1) A_m\over (\mu+1) C} \right] \nonumber \\
&& +{(\mu-1)(A_m+B) \over 2 A_m} \arctan \left[ {(\mu-1) B-(\nu-1) A_m\over (\mu-1) C} \right] \nonumber \\
&& -{(\mu-1)(A_m-B) \over 2 A_m} \arctan \left[ {(\mu-1) B-(\nu+1) A_m\over (\mu-1) C} \right] \nonumber \\
&& +{(\nu+1)(A_p-B) \over 2 A_p} \arctan \left[ {(\nu+1) B-(\mu-1) A_p\over (\nu+1) C} \right] \nonumber \\
&& -{(\nu+1)(A_p+B) \over 2 A_p} \arctan \left[ {(\nu+1) B-(\mu+1) A_p\over (\nu+1) C} \right] \nonumber \\
&& +{(\nu-1)(A_p+B) \over 2 A_p} \arctan \left[ {(\nu-1) B-(\mu-1) A_p\over (\nu-1) C} \right] \nonumber \\
&& -{(\nu-1)(A_p-B) \over 2 A_p} \arctan \left[ {(\nu-1) B-(\mu+1) A_p\over (\nu-1) C} \right] \nonumber \\
&& -{(\mu+1) B\over A_m} \arctan \left[ {\nu A_m-(\mu+1) B\over (\mu+1) C} \right]
   +{(\mu-1) B\over A_m} \arctan \left[ {\nu A_m-(\mu-1) B\over (\mu-1) C} \right] \nonumber \\
&& -{(\nu+1) B\over A_p} \arctan \left[ {\mu A_p-(\nu+1) B\over (\nu+1) C} \right]
   +{(\nu-1) B\over A_p} \arctan \left[ {\mu A_p-(\nu-1) B\over (\nu-1) C} \right] \nonumber \\
&& +\mu \arctan \left[ {\mu B-(\nu+1) A_m\over \mu C} \right]
   -\mu \arctan \left[ {\mu B-(\nu-1) A_m\over \mu C} \right] \nonumber \\
&& +\nu \arctan \left[ {\nu B-(\mu+1) A_p\over \nu C} \right]
   -\nu \arctan \left[ {\nu B-(\mu-1) A_p\over \nu C} \right] \nonumber \\
&& +{(\mu+1)C \over 4 A_m} \left \{
   \ln \left[ 2 (\mu+1) (\nu-1) B-(\mu+1)^2 A_p-(\nu-1)^2 A_m \right] \right. \nonumber \\
&& +\ln \left[ 2 (\mu+1) (\nu+1) B-(\mu+1)^2 A_p-(\nu+1)^2 A_m \right] \nonumber \\
&& \left. -2 \ln \left[ 2 (\mu+1) \nu B-(\mu+1)^2 A_p-\nu^2 A_m \right] \right \} \nonumber \\
&& -{(\mu-1)C \over 4 A_m} \left \{
   \ln \left[ 2 (\mu-1) (\nu+1) B-(\mu-1)^2 A_p-(\nu+1)^2 A_m \right] \right. \nonumber \\
&& +\ln \left[ 2 (\mu-1) (\nu-1) B-(\mu-1)^2 A_p-(\nu-1)^2 A_m \right] \nonumber \\
&& \left. -2 \ln \left[ 2 (\mu-1) \nu B-(\mu-1)^2 A_p-\nu^2 A_m \right] \right \} \nonumber \\
&& +{(\nu+1)C \over 4 A_p} \left \{
   \ln \left[ 2 (\mu+1) (\nu+1) B-(\mu+1)^2 A_p-(\nu+1)^2 A_m \right] \right. \nonumber \\
&& +\ln \left[ 2 (\mu-1) (\nu+1) B-(\mu-1)^2 A_p-(\nu+1)^2 A_m \right] \nonumber \\
&& \left. -2 \ln \left[ 2 \mu (\nu+1) B-\mu^2 A_p-(\nu+1)^2 A_m \right] \right \} \nonumber \\
&& -{(\nu-1)C \over 4 A_p} \left \{
   \ln \left[ 2 (\mu-1) (\nu-1) B-(\mu-1)^2 A_p-(\nu-1)^2 A_m \right] \right. \nonumber \\
&& +\ln \left[ 2 (\mu+1) (\nu-1) B-(\mu+1)^2 A_p-(\nu-1)^2 A_m \right] \nonumber \\
&& \left. -2 \ln \left[ 2 \mu (\nu-1) B-\mu^2 A_p-(\nu-1)^2 A_m \right] \right \} ,
\label{volledig}
\end{eqnarray}
where $A_p=1+A$ and $A_m=1-A$, with $A$, $B$ and $C$ as defined in Eqs.~(\ref{parma}-\ref{parmc}).

\newpage
\bibliography{References}

\newpage
\section*{List of Figure Captions}

\begin{figure}[htbp]
\centering
\includegraphics[width=9cm]{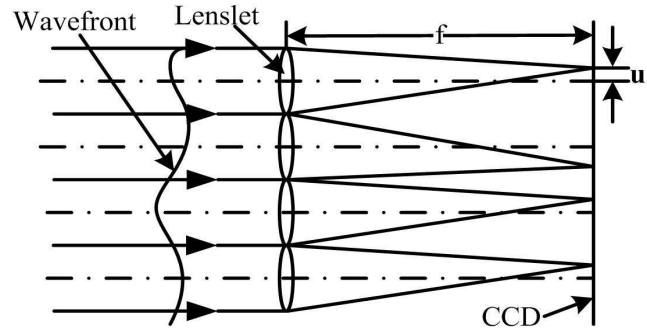}
\caption{One-dimensional representation of a Shack-Hartmann wavefront sensor. ChenFig1.eps.} \label{dia}
\end{figure}

Fig.~\ref{dia} One-dimensional representation of a Shack-Hartmann wavefront sensor.

\begin{figure}[htbp]
\centering
\includegraphics[width=9cm]{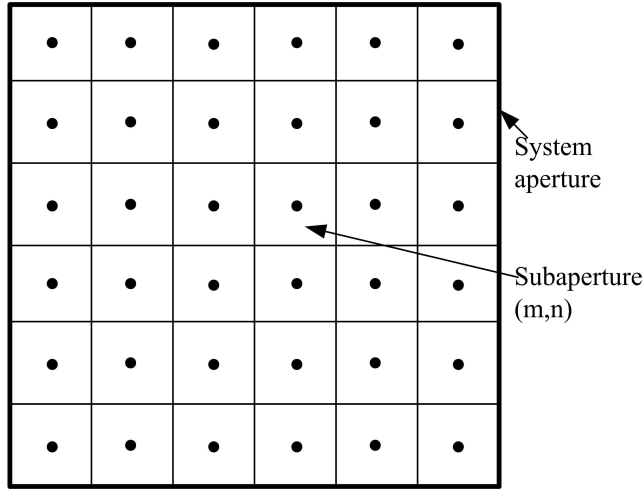}
\caption{An array of subapertures (small squares) within the system aperture of the SHWS. The average phase slope values are associated with the dots inside the small squares. ChenFig2.eps.} \label{array}
\end{figure}

Fig.~\ref{array} An array of subapertures (small squares) within the system aperture of the SHWS. The average phase slope values are associated with the dots inside the small squares.

\begin{figure}[htbp]
\centering
\includegraphics[width=9cm]{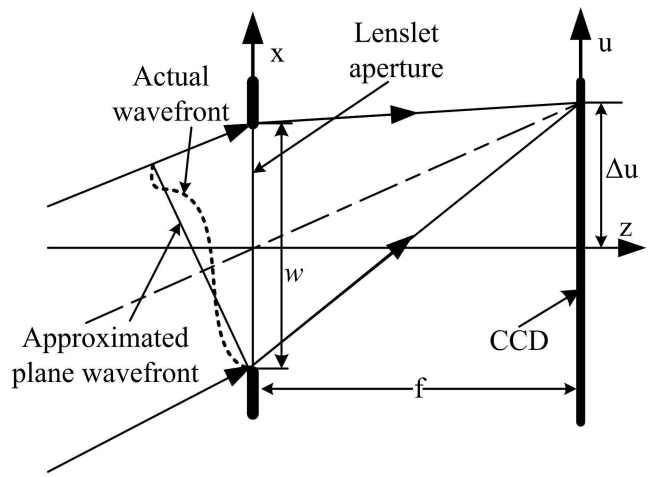}
\caption{One-dimensional representation of one lenslet in the SHWS, showing the shift of the focal point due to the average tilt of the incident wavefront. ChenFig3.eps.}
\label{ft}
\end{figure}

Fig.~\ref{ft} One-dimensional representation of one lenslet in the SHWS, showing the shift of the focal point due to the average tilt of the incident wavefront.

\begin{figure}[htbp]
\centering
\includegraphics{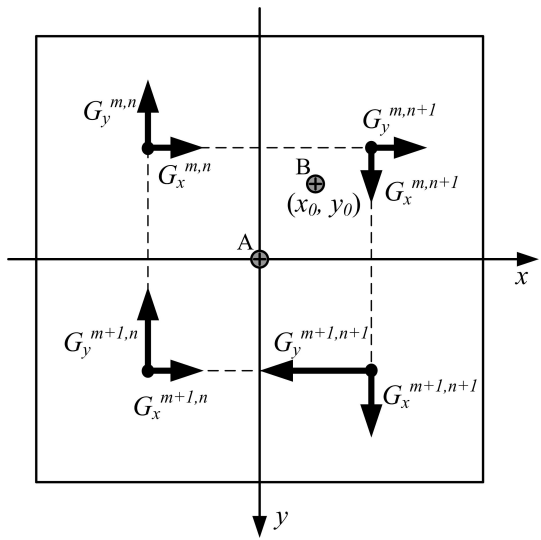}
\caption{Circulation $D^{m,n}$ over four subapertures with a singularity located either at the center (assumed to be the origin), denoted by {\bf A} or at some arbitrary location $(x_0,y_0)$ denoted by {\bf B}.  The four subapertures are represented by the four squares. The dot at the center of each subaperture is the position with which the average phase slope value ${\bf G}$ of that subaperture is associated. The arrows represent the components of ${\bf G}$. The dashed-lines represent the contour used for the calculation of the circulation. ChenFig4.eps.}
\label{foursub}
\end{figure}

Fig.~\ref{foursub} Circulation $D^{m,n}$ over four subapertures with a singularity located either at the center (assumed to be the origin), denoted by {\bf A} or at some arbitrary location $(x_0,y_0)$ denoted by {\bf B}.  The four subapertures are represented by the four squares. The dot at the center of each subaperture is the position with which the average phase slope value ${\bf G}$ of that subaperture is associated. The arrows represent the components of ${\bf G}$. The dashed-lines represent the contour used for calculation of the circulation.

\begin{figure}[htbp]
\centering
\includegraphics{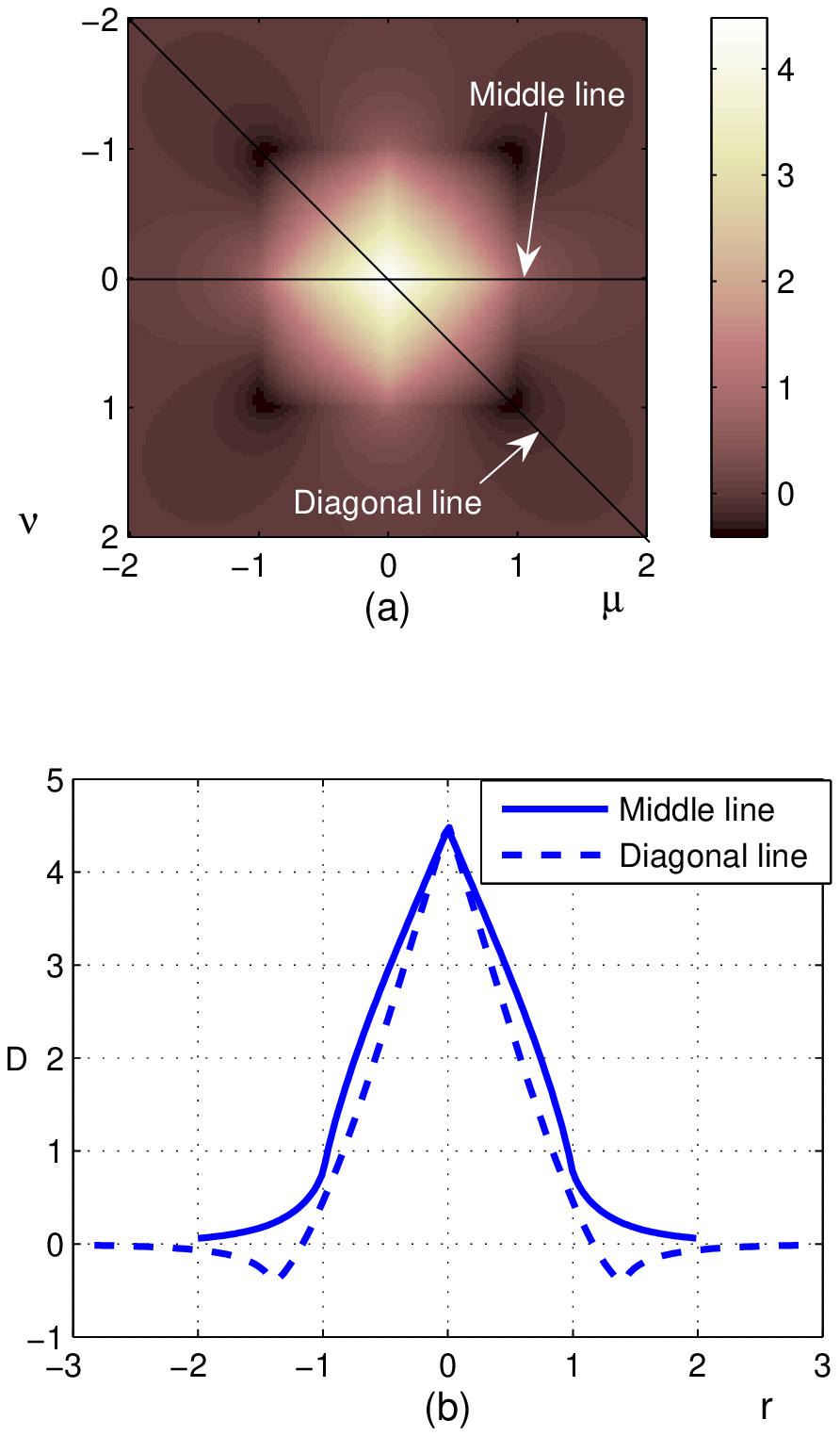}
\caption{(Color online) Circulation $D$ for a canonical singularity. A topview of $D$ is shown in (a) as a function of the relative position of the singularity inside the four subaperture area, shown in Fig.~\ref{foursub}, for $-2<\mu<2$ and $-2<\nu<2$. One-dimensional functions of $D$ are plotted as functions of $r=\sqrt{\mu^2+\nu^2}$ in (b) along the `diagonal line' and `middle line,' respectively, as indicated in (a). ChenFig5.eps.}
\label{canshape}
\end{figure}

Fig.~\ref{canshape} Circulation $D$ for a canonical singularity. A topview of $D$ is shown in (a) as a function of the relative position of the singularity inside the four subaperture area, shown in Fig.~\ref{foursub}, for $-2<\mu<2$ and $-2<\nu<2$. One-dimensional functions of $D$ are plotted as functions of $r=\sqrt{\mu^2+\nu^2}$ in (b) along the `diagonal line' and `middle line,' respectively, as indicated in (a).

\begin{figure}[htbp]
\centering
\includegraphics{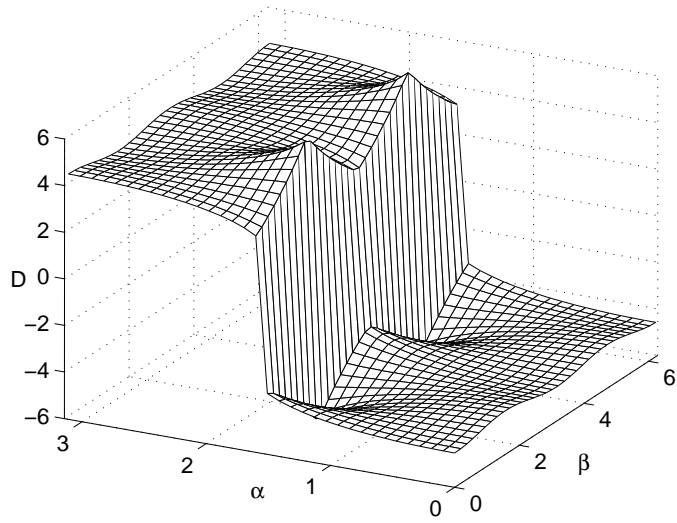}
\caption{Peak value of the circulation $D$ as a function of the morphology angles $0<\alpha<\pi$ and $0<\beta<2\pi$. The jump at $\alpha=\pi/2$ is due to the change of the topological charge of the singularity. ChenFig6.eps.}
\label{peak}
\end{figure}

Fig.~\ref{peak} Peak value of the circulation $D$ as a function of the morphology angles $0<\alpha<\pi$ and $0<\beta<2\pi$. The jump at $\alpha=\pi/2$ is due to the change of the topological charge of the singularity.

\begin{figure}[htbp]
\centering
\includegraphics{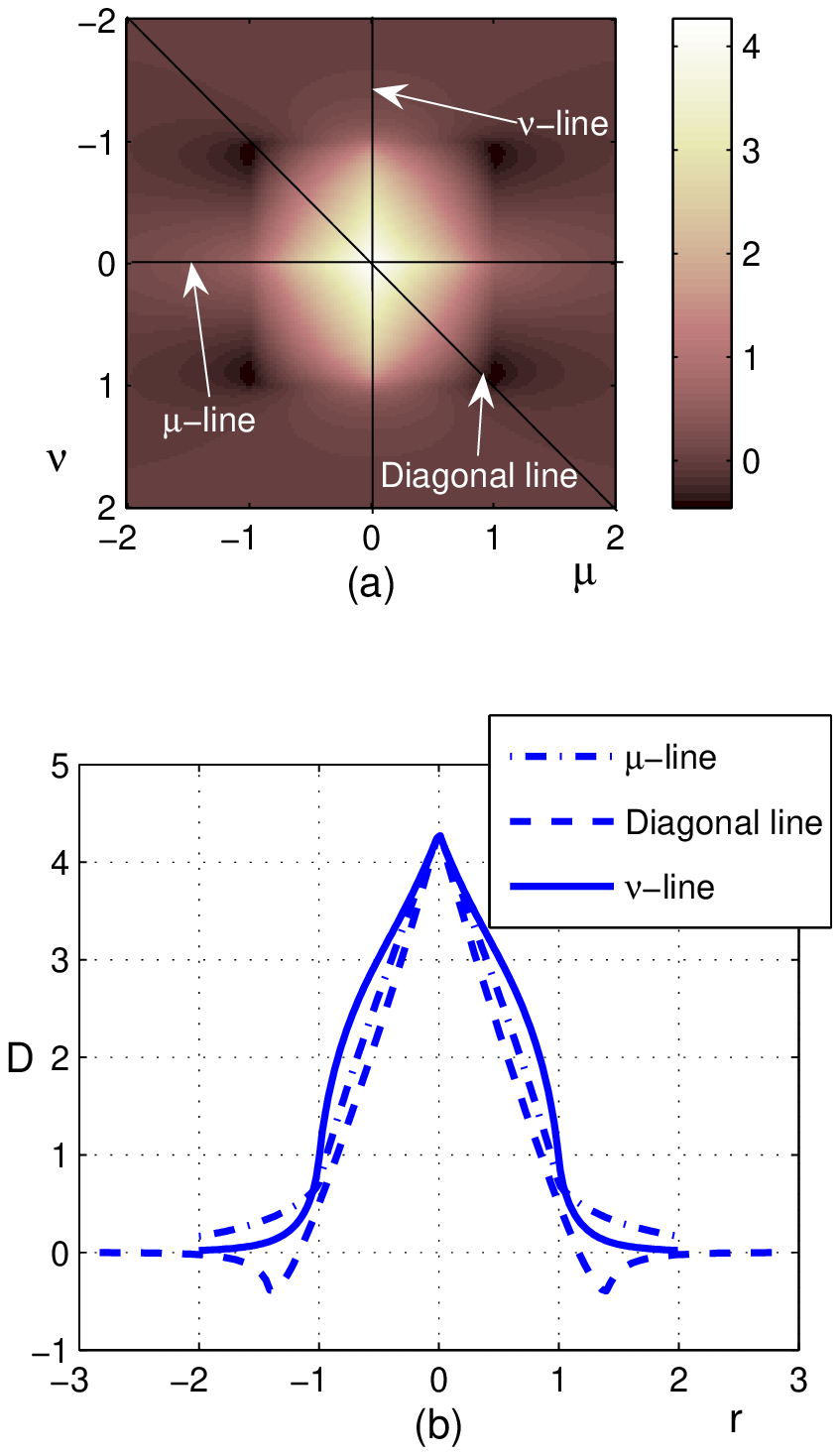}
\caption{(Color online) Circulation $D$ for a noncanonical singularity, with $\alpha=\pi/4$ and $\beta=\pi$. A topview of $D$ is shown in (a) as a function of the relative position of the singularity inside the four subaperture area, shown in Fig.~\ref{foursub}, for $-2<\mu<2$ and $-2<\nu<2$. One-dimensional functions of $D$ are plotted as functions of $r=\sqrt{\mu^2+\nu^2}$ in (b) along the `diagonal line', `$\mu$-line' and `$\nu$-line,' respectively, as indicated in (a). ChenFig7.eps.}
\label{ex1}
\end{figure}

Fig.~\ref{ex1} Circulation $D$ for a noncanonical singularity, with $\alpha=\pi/4$ and $\beta=\pi$. A topview of $D$ is shown in (a) as a function of the relative position of the singularity inside the four subaperture area, shown in Fig.~\ref{foursub}, for $-2<\mu<2$ and $-2<\nu<2$. One-dimensional functions of $D$ are plotted as functions of $r=\sqrt{\mu^2+\nu^2}$ in (b) along the `diagonal line', `$\mu$-line' and `$\nu$-line,' respectively, as indicated in (a).

\begin{figure}[htbp]
\centering
\includegraphics{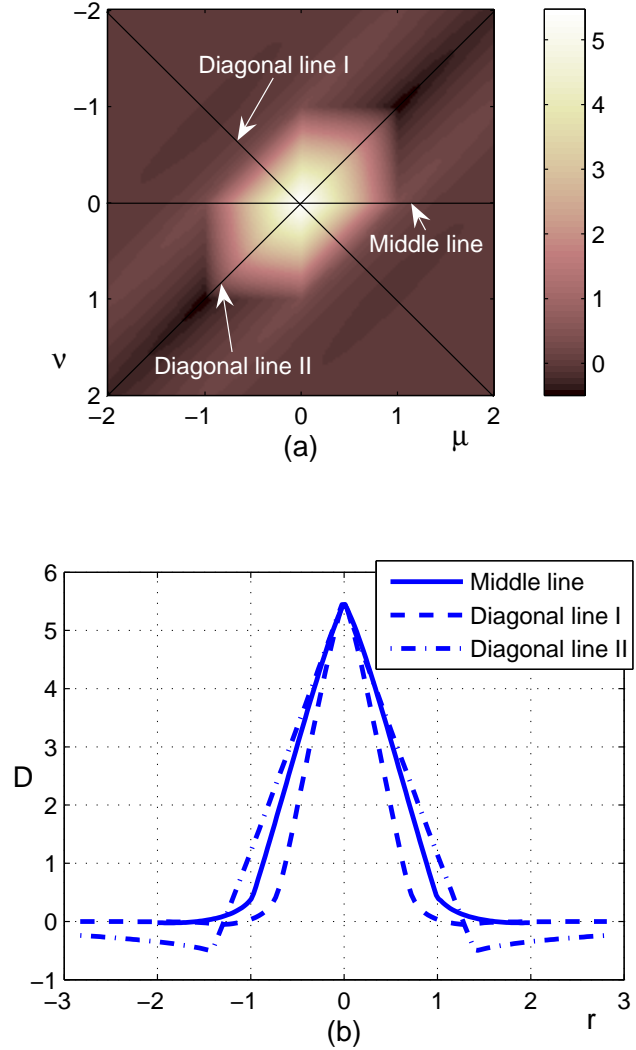}
\caption{(Color online) Circulation $D$ for a noncanonical singularity, with $\alpha=4\pi/9$ and $\beta=\pi/2$. A topview of $D$ is shown in (a) as a function of the relative position of the singularity inside the four subaperture area, shown in Fig.~\ref{foursub}, for $-2<\mu<2$ and $-2<\nu<2$. One-dimensional functions of $D$ are plotted as functions of $r=\sqrt{\mu^2+\nu^2}$ in (b) along the `diagonal line I' (perpendicular to the orientation of the singularity), `diagonal line II' (along the orientation of the singularity) and `middle line,' respectively, as indicated in (a). ChenFig8.eps.}
\label{ex2}
\end{figure}

Fig.~\ref{ex2} Circulation $D$ for a noncanonical singularity, with $\alpha=4\pi/9$ and $\beta=\pi/2$. A topview of $D$ is shown in (a) as a function of the relative position of the singularity inside the four subaperture area, shown in Fig.~\ref{foursub}, for $-2<\mu<2$ and $-2<\nu<2$. One-dimensional functions of $D$ are plotted as functions of $r=\sqrt{\mu^2+\nu^2}$ in (b) along the `diagonal line I' (perpendicular to the orientation of the singularity), `diagonal line II' (along the orientation of the singularity) and `middle line,' respectively, as indicated in (a).

\begin{figure}[htbp]
\centering
\includegraphics{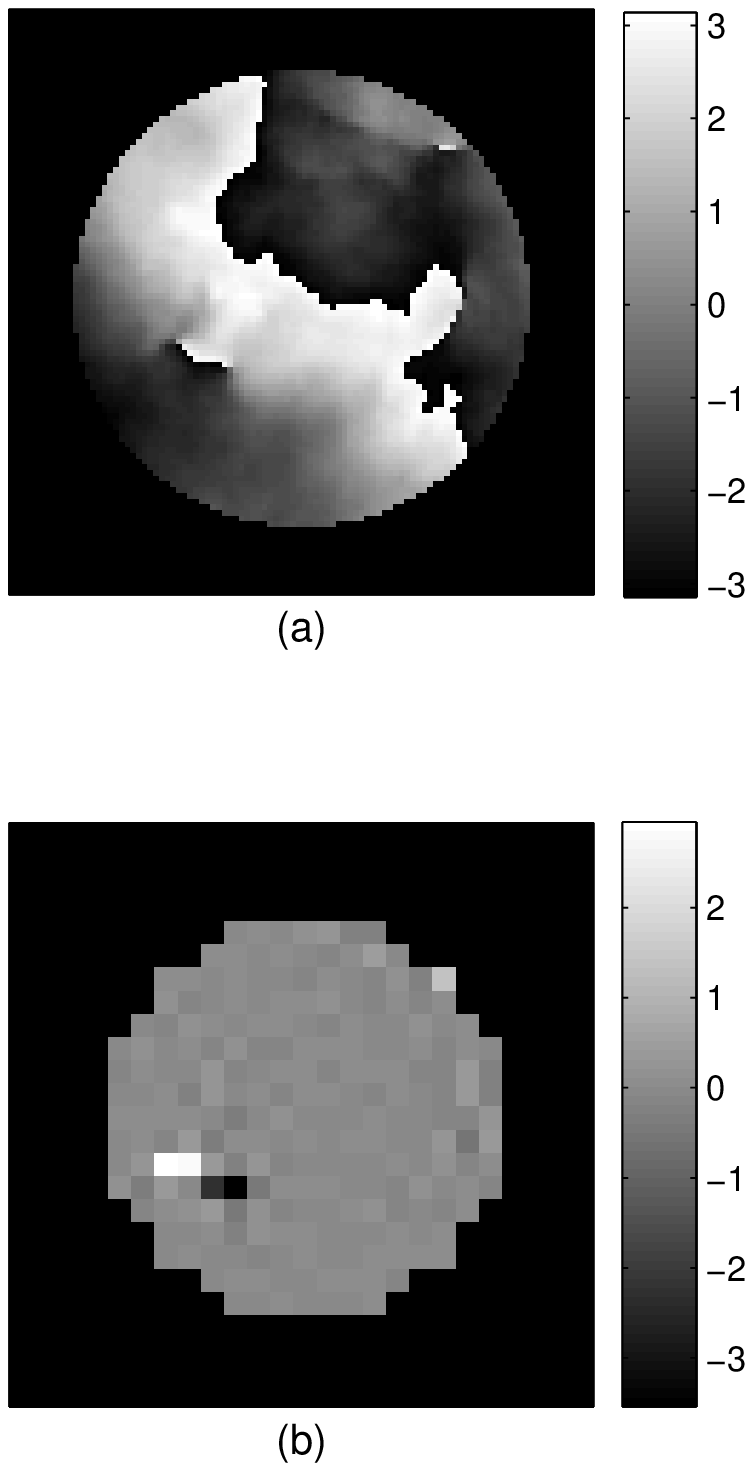}
\caption{Numerical simulation results for a Gaussian beam that propagated over a distance of 100~km through a turbulent atmosphere. The resulting phase of the beam inside the system aperture is shown in (a). There are two pairs of oppositely charged phase singularities. The pairs are, respectively, located at the lower left and the upper right of the system aperture. The circulation $D$, numerically calculated from the output of the Shack-Hartmann wavefront sensor, is shown in (b). ChenFig9.eps.}
\label{numfig}
\end{figure}

Fig.~\ref{numfig} Numerical simulation results for a Gaussian beam that propagated over a distance of 100~km through a turbulent atmosphere. The resulting phase of the beam inside the system aperture is shown in (a). There are two pairs of oppositely charged phase singularities. The pairs are, respectively, located at the lower left and the upper right of the system aperture. The circulation $D$, numerically calculated from the output of the Shack-Hartmann wavefront sensor, is shown in (b).


\newpage

\newpage

\newpage

\newpage

\newpage

\newpage

\newpage

\newpage

\newpage

\end{document}